\documentclass[]{spie}  %>>> use for US letter paper
\usepackage{amsmath,amsfonts,amssymb}
\usepackage{gensymb}
\usepackage{graphicx}
\usepackage[colorlinks=true, allcolors=blue]{hyperref}

\usepackage{wrapfig}

\title{Multi-Source Static CT with Adaptive Fluence Modulation to Minimize Hallucinations in Generative Reconstructions}

\author[a]{Matthew Tivnan}
\author[a]{Amar Gupta}
\author[a]{Kai Yang}
\author[a]{Dufan Wu}
\author[a]{Rajiv Gupta}
\affil[a]{Massachusetts General Hospital and Harvard Medical School, 55 Fruit St., Boston, MA, USA}

\begin{document} 
\maketitle
% \vspace{-6mm}
\begin{abstract}
Multi-source static Computed Tomography (CT) systems have introduced novel opportunities for adaptive imaging techniques. This work presents an innovative method of fluence field modulation using spotlight collimators. These instruments block positive or negative fan angles of even and odd indexed sources, respectively. Spotlight collimators enable volume of interest imaging by increasing relative exposure for the overlapping views. To achieve high quality reconstructions from sparse-view low-dose data, we introduce a generative reconstruction algorithm called Langevin Posterior Sampling (LPS), which uses a score based diffusion prior and physics based likelihood model to sample a posterior random walk. We conduct simulation-based experiments of head CT imaging for stroke detection and we demonstrate that spotlight collimators can effectively reduce the standard deviation and worst-case scenario hallucinations in reconstructed images. Compared to uniform fluence, our approach shows a significant reduction in posterior standard deviation. This highlights the  potential for spotlight collimators and generative reconstructions to improve image quality and diagnostic accuracy of multi-source static CT. 
\end{abstract}

% Include a list of keywords after the abstract 
\keywords{Static CT,  Multi-Source CT, Fluence Field Modulation, Diffusion Models, Hallucinations }

% \vspace{-4mm}

\section{INTRODUCTION}

Multi-source static Computed Tomography (CT) is an emerging technology that uses an array of x-ray sources to achieve projections for different view angles rather than a single source on a rotating gantry \cite{billingsley2023clinical,wang2024iterative, yao2021novel, cramer2018stationary}. In practice, this is made possible through the use of carbon nanotube x-ray sources which are inexpensive and compact. The advantages of static CT include light weight, robust mechanical design, lower duty cycle per source for heat dissipation, fast rotation speeds through electronic switching, and more. In this work, we present a novel method of fluence field modulation that is only compatible with static CT. Our proposed instrumentation has no moving parts, but is capable of delivering higher fluence to specific parts of the object. Through simulation-based experiments, we demonstrate our method has high potential for applications such as volume of interest imaging and reducing hallucinations in generative reconstructions.

% \begin{wrapfigure}[19]{l}{0.5\textwidth}
\begin{figure}
    % \vspace{-8mm}
    \begin{center}
    \includegraphics[width=0.75\textwidth]{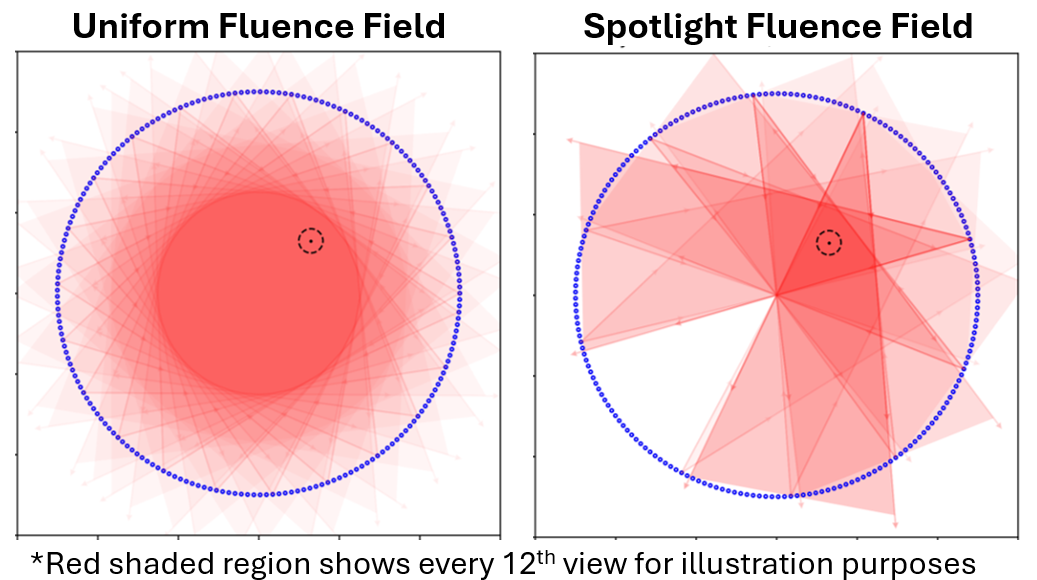}
  \end{center}
  \caption{Illustration of a multi-source static computed tomography system. For the uniform fluence field, each source has fan angle $\pm 30^{\circ}$. For spotlight collimated system, the even and odd index sources illuminate positive and negative fan angles, respectively. By modulating the current of each source, one can design the spotlight fluence field to have higher effective exposure for a volume of interest. }
  \label{fig:spotlight_cartoon.png}
\end{figure}
% \end{wrapfigure}

There are several existing methods for fluence modulated CT imaging. One technique is tube current modulation (TCM) which involves increasing fluence for the lateral projections and decreasing fluence for the anteroposterior projections with the goal of uniform image quality \cite{kalra2004techniques}. Another option, Multiple Aperture Devices (MADs), consist of a structured pattern of apertures and use robotic translation to modulate fluence fields for volume of interest imaging \cite{stayman2016fluence,wang2019volume}. Another technique called a Digital Beam Attenuator (DBA) uses control of the overlap of iron edges that change the effective filter thickness  \cite{szczykutowicz2014experimental,szczykutowicz2013design, szczykutowicz2013design2}. It is also possible to use a multi-leaf collimator (MLC) such as those used in intensity modulated radiation therapy \cite{szczykutowicz2015realization}. Except for TCM, these devices all require dynamic instrumentation with robotic control because they need to change the fluence for a single x-ray source on a rotating gantry. Our proposed design enables fluence field modulation with no moving parts.

Our proposal is to introduce \emph{spotlight collimators} into a multi-source static CT system, which block out a different part of the x-ray beam for each source. The simplest example is shown in Figure 1, where the even index sources are collimated to pass positive fan angles only and odd index sources are collimated to pass negative fan angles only. While it may seem counterintuitive to remove some lines of response from the system, the benefit comes from the ability to selectively illuminate certain regions of the patient with higher effective exposure. In this work, we use simulations to investigate the potential advantages of fluence modulation with spotlight collimators. To normalize radiation dose with uniform fluence, we use double exposure with spotlight collimators. 

Due to the view sparsity of realistic multi-source static CT systems and the increased sparsity introduced by spotlight collimators, the image reconstruction algorithm must take advantage of strong priors. In this work, we use a generative image reconstruction algorithm; specifically, we use Langevin posterior sampling with a physics-based measurement likelihood term and a probabilistic prior term from a pre-trained score-based diffusion model. Our results show that the model \emph{hallucinates} when there is little information in the measurements. That is, many possible anatomical structures could plausibly underlie the noisy measurements, and so the variance of the posterior random walk is high. Through our simulation-based experiments, we show that spotlight collimators can be used to apply higher exposure to areas with more hallucinations, resulting in higher image quality.

% \vspace{-4mm}

\section{METHODS}

\subsection{Simulating Fluence Field Modulation with Spotlight Collimators in Static CT}

We used a nonlinear forward model to simulate noise and measurements as shown below:

% \vspace{-4mm}

\begin{equation}
\mathbf{\bar{y}(\mathbf{x})} = \mathbf{I_0} \circ \exp{\Big(-\mathbf{A}{\mathbf{x}}\Big)}
\end{equation}

% \vspace{-4mm}

\noindent where $\mathbf{x}\in \Re^{(N\times1)}$ is a flattened vector representation of the ground truth image of attenuation coefficients assuming a 60keV monoenergetic source spectrum, $\mathbf{A}\in \Re^{(M\times N)}$ is a forward projection matrix, $\mathbf{I_0}\in\Re^{(M\times1)}$ is the number of photons per line of response, and $\mathbf{\bar{y}(\mathbf{x})}\in\Re^{(M\times1)}$ is the expected photon counts. 

% \begin{wrapfigure}[22]{l}{0.5\textwidth}
%     \vspace{-8mm}
%     \begin{center}
%     \includegraphics[width=0.48\textwidth]{figures/spotlight_fluence_sinogram.png}
%   \end{center}
%   \caption{Illustration of a multi-source static computed tomography system. For the uniform fluence field, each source has fan angle $\pm 30^{\circ}$. For spotlight collimated system, the even and odd index sources illuminate positive and negative fan angles, respectively. By modulating the current of each source, one can design the spotlight fluence field to have higher effective exposure for a volume of interest. }
%   \label{fig:spotlight_fluence_sinogram}
% \end{wrapfigure} 
The $\circ$ symbol indicates elementwise multiplication and the exponential operation is elementwise. We consider the measurements to follow an independent Poisson distribution $\mathbf{y}|\mathbf{x} \sim \mathcal{P}(\mathbf{\bar{y}}(\mathbf{x}))$. 

To model the spotlight collimators, we modified $\mathbf{A}$ by deleting the matrix rows corresponding to lines of response that are blocked by the collimator. In terms of practical implementation, this is accomplished by forward/back projecting even and odd views together and limiting the fan angle appropriately, followed by an operation to stitch the data together in a common projection domain. 

For fluence modulation, the exposure of each source, as modeled by $\mathbf{I_0}$, should be increased and decreased where needed for the given imaging task. As shown in Figure \ref{fig:spotlight_fluence_sinogram}, volume of interest (VOI) imaging can be accomplished by forward-projecting the volume to get a sinogram domain mask, and then reducing current on the sources that are not overlapping the VOI. After normalizing by total exposure, the effect will be higher fluence in the volume of interest relative to a uniform fluence scan. 

% \vspace{-2mm}

\subsection{Generative Reconstruction Algorithm: Langevin Posterior Sampling}

We implemented diffusion posterior sampling (DPS) \cite{chung2022diffusion} with simulated static CT measurements following the methods of previous work \cite{li2024diffusion,lopez2024stationary}.
We used the variance-exploding process \cite{song2020score} as shown below.

% \vspace{-4mm}

\begin{equation}
\mathbf{dx}_t = g \mathbf{dw}_t
\end{equation}

% \vspace{-6mm}

\noindent and to sample from the posterior with DPS we can use the reverse process.

% \vspace{-8mm}

\begin{gather}
    \mathbf{dx}_t = g^2 \nabla_{\mathbf{x}_t} \log p_t(\mathbf{x}_t|\mathbf{y})dt  + g \mathbf{dw}_t \\
    \mathbf{dx}_t \approx g^2  [\nabla_{\mathbf{x}_t}\log p_t(\mathbf{x}_t) + \gamma\nabla_{\mathbf{x}}\log p(\mathbf{y} |\mathbf{x} = \mathbf{x}_t)]dt  + g \mathbf{dw}_t
    \label{eq:DPS}
\end{gather}
% \vspace{-8mm}

\noindent where $\nabla_{\mathbf{x}_t}\log p_t(\mathbf{x}_t)$ is a prior score function which is approximated with a neural network and $\nabla_{\mathbf{x}}\log p(\mathbf{y} |\mathbf{x})$ is the likelihood score function. For our independent Poisson measurements, the score function is given by
% \vspace{-6mm}

\begin{equation}
\nabla_{\mathbf{x}}\log p(\mathbf{y} |\mathbf{x}) = \mathbf{A}^T (\mathbf{\bar{y}}(\mathbf{x}) - \mathbf{y}) 
\end{equation}
% \vspace{-6mm}

It is important for our methods in the following section to generate many samples from the posterior to quantify the uncertainty of the generative model, commonly referred to as hallucinations. To that end, we used the annealed Langevin sampling algorithm described in a previous work \cite{song2019generative}. We used the same posterior score function previously derived in \eqref{eq:DPS} to implement a method we refer to as \emph{Langevin Posterior Sampling} (LPS) using the stochastic differential equation below.
% \vspace{-4mm}

\begin{equation}
    \mathbf{dx}_s = \frac{1}{2} \alpha^2  [\nabla_{\mathbf{x}_t}\log p_t(\mathbf{x}_t = \mathbf{x}_s) + \nabla_{\mathbf{x}}\log p(\mathbf{y} |\mathbf{x} = \mathbf{x}_s)]dt  + \alpha \hspace{1mm} \mathbf{dw}_s
\end{equation}
% \vspace{-6mm}

The form of this equation is similar to reverse diffusion sampling but with a factor of $\frac{1}{2}$ in front of the score term. As a result, the distribution will not change over time but the samples will be a random walk on the posterior distribution. Notice in the LPS stochastic differential equation we have removed the coefficient $\gamma$. This is because we set $\gamma=1$ for principled Bayesian inference. That is, we are setting the proper weights on the prior and likelihood terms such that the random walk is truly over the posterior distribution. In many other works using DPS, they set this coefficient to less than one so the step size can be increased without numerical instability. We were able to achieve $\gamma=1$ by reducing the step size $\alpha$ and running many iterations. This stochastic process results in a random walk on the posterior distribution defined by a diffusion-based prior score function and a physics model-based likelihood score function. We do not claim that this random walk will generate higher quality reconstructions than a single sample of DPS. Rather, it allows for Bayesian estimation by generating many samples from the posterior distribution of high-quality images given low-quality measurements. By taking the standard deviation across realizations of the random walk, we can characterize the magnitude and position of hallucinations in the generative reconstructions.

% \vspace{-2mm}

\subsection{Experiment: Adaptive Fluence Field Modulation to Minimize Hallucinations}

We conducted a simulation-based experiment to evaluate whether fluence modulation with spotlight collimators could be used to reduce hallucinations in generative reconstructions. We simulated a static CT system with a full ring of 72 sources with 400 mm source-to-axis distance and 400 mm source-to-detector distance using the LEAP cone beam forward projector \cite{kim2023differentiable}. We consider a scenario where 10\% exposure is used for a low-dose scout CT using uniform fluence on all sources (with spotlight collimators). Then we run the LPS algorithm to sample a random walk on the posterior distribution of head CT images given the low-dose scout measurements. The regions of greatest uncertainty will be subject to the greatest variability. We are particularly interested in hallucinations in the brain tissue which is critical for diagnosing many pathologies, so we apply a threshold-based bone mask to the standard deviation map to create our hallucination map. Then, we forward projected this hallucination map to the projection domain to generate a fluence plan for the remaining 90\% exposure. For now, we set the fluence proportional to the magnitude of the forward projected hallucination map. Then we run the LPS algorithm again conditioned on the measurements from both the scout and final scans.  For a comparison case, we also simulate uniform fluence with no spotlight collimators and normalized exposure to $10^6$ photons per pixel per view on average for both cases.

For training data, we extracted 10,800 axial head CT images with shape $256\times256$ from Task 1 of the SynthRAD2023 dataset \cite{thummerer2023synthrad2023}. We used 10,000 images for training and 800 for evaluation. There is some noise and blur in these images, but it is much higher image quality than what is possible with low-dose sparse view CT simulated in this work, so we treat it as ground truth.  For the diffusion backbone network, we used HuggingFace conditional U-Net \cite{ronneberger2015u}. We trained the prior score estimator with 10,000 epochs with 1,000 iterations per epoch and 16 full $256\times256$ images. To initialize LPS, we ran DPS with 256 evenly spaced time steps using a second order Heun sampler \cite{karras2022elucidating}. Then, we ran 1024 iterations of LPS, setting the step size dynamically such that the magnitude of the score term is always 5 HU on average across all pixels.  To evaluate the resulting image distributions, we computed root mean squared error (RMSE), root mean squared bias (RMS bias), and standard deviation of the posterior samples for both spotlight fluence and uniform fluence cases. We also evaluated the $90^{\text{th}}$ percentile standard deviation across all pixels of the standard deviation map to evaluate the worst-case-scenario hallucinations. For all of these metrics, we used a brain mask (0 to 80 HU) to evaluate only the errors in the brain region to avoid the bone regions dominating the error metrics. We repeated this process for 800 evaluations and report the mean and standard deviation of each error metric over the patient population.

% \vspace{-4mm}

\begin{figure}
    \centering
    \includegraphics[width=0.99\linewidth]{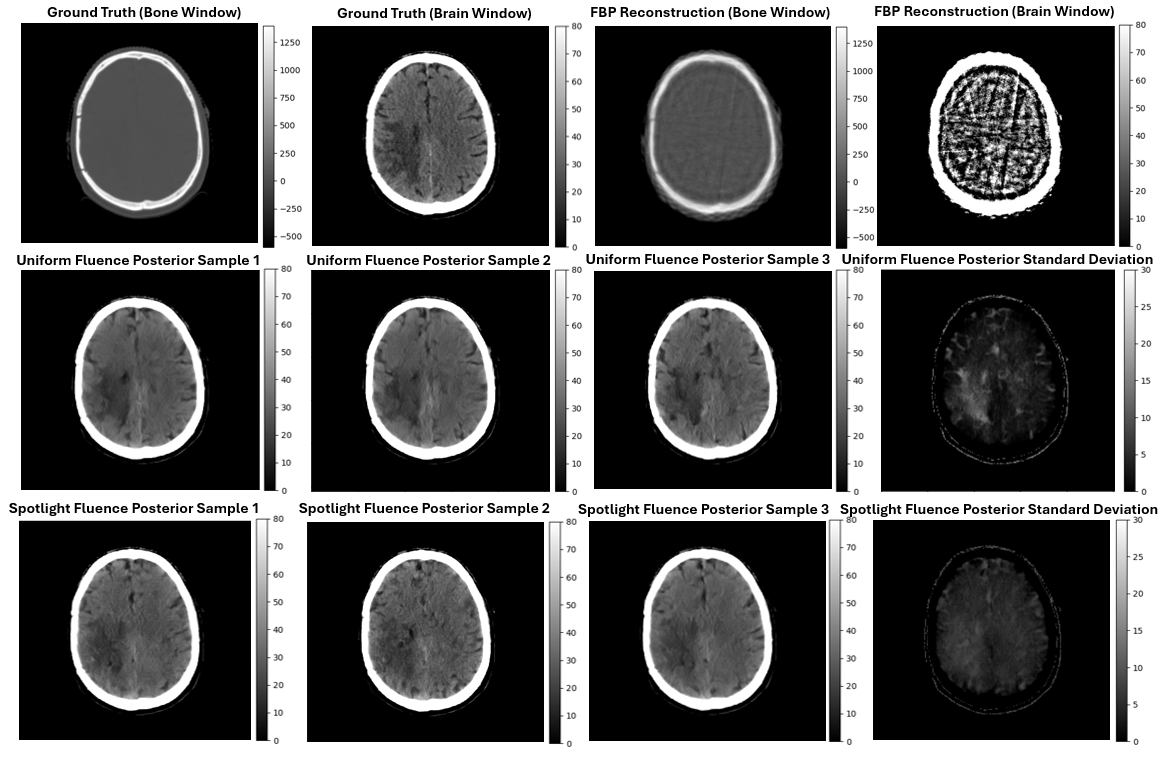}
    \caption{Image results showing hallucinations are reduced using spotlight fluence. The top row shows the ground truth image in a bone and brain window as well as FBP based reconstructions from simulated multi-sources static CT data (with uniform fluence). The second row shows three posterior samples from the LPS algorithm as well as a posterior standard deviation map for the system using uniform fluence. The third row shows the spotlight fluence scout with 10\% exposure used for fluence planning. The fourth row shows the results for optimized spotlight fluence. Total hallucination magnitude is somewhat reduced and the worse-case-scenario hallucination magnitude is greatly reduced. }
    \label{fig:enter-label}
\end{figure}

\section{RESULTS}

% \vspace{-2mm}

Our results demonstrate a significant reduction in hallucinations when using the proposed spotlight fluence modulation compared to uniform fluence. As shown in Table \ref{tab:results}, the standard deviation across the posterior samples is lower for the spotlight fluence case, indicating reduced variability in the reconstructed images. Reductions in RMSE and RMS Bias were not found to be statistically significant in this study but the p-value for RMSE is close to significance at a p-value of 0.0547. The most dramatic improvement is over the 90th percentile standard deviation, a measure of worst-case scenario hallucinations, is also significantly reduced, highlighting the effectiveness of the spotlight fluence method.
% \vspace{-4mm}

\begin{table}[h!]
\centering
\footnotesize
\caption{Comparison of Error Metrics for Spotlight Fluence and Uniform Fluence}
\label{tab:results}
\begin{tabular}{|l|c|c|c|}
\hline
\textbf{Error Metric} & \textbf{Uniform Fluence} & \textbf{Spotlight Fluence} & \textbf{p-value (n=800)} \\ \hline
\textbf{RMSE (HU)} & $11.4 \pm 5.1$ & $10.9 \pm 5.3$ & $5.47 \times 10^{-2}$ \\ \hline
\textbf{RMS Bias (HU)} & $4.9 \pm 2.5$ & $4.8 \pm 2.7$ & $4.42 \times 10^{-1}$ \\ \hline
\textbf{Standard Deviation (HU)} & $10.4 \pm 3.3$ & $9.6 \pm 3.6$ & $3.89 \times 10^{-6}$ \\ \hline
\textbf{90th Percentile Std. Dev. (HU)} & $28.3 \pm 4.7$ & $20.1 \pm 3.8$ & $6.80 \times 10^{-229}$ \\ \hline
\end{tabular}
\end{table}

% \vspace{-7mm}

\section{CONCLUSION}

% \vspace{-2mm}

In this work, we introduced a novel approach to fluence modulation in multi-source static CT systems using spotlight collimators. We also introduced the LPS algorithm for Bayesian characterization of the posterior associated with a trained prior. One limitation of our work is the lack of physical data. In the future, we plan to test spotlight collimators and LPS reconstruction on physical measurements from a prototype  multi-source static CT system. The spotlight fluence method demonstrated a significant reduction in posterior standard deviation compared to uniform fluence. These findings suggest that adaptive fluence modulation has the potential to improve multi-source static CT image quality. 

\newpage

% References
\bibliography{report} % bibliography data in report.bib
\bibliographystyle{spiebib} % makes bibtex use spiebib.bst

\end{document}